\documentclass{appolb}
\usepackage{axodraw,epsfig,cite,amsmath,amssymb,amsfonts}

\input{paperdef}
\graphicspath{{figs/}}

\begin{document}

\thispagestyle{empty}
\setcounter{page}{0}
\def\thefootnote{\fnsymbol{footnote}}

\begin{flushright}
\mbox{}
\end{flushright}

\vspace{1cm}

\begin{center}

{\large\sc {\bf Higgs Physics at the LHC: Some Theory Aspects}}
\footnote{lecture given at the {\em Spanish Winter Meeting 2008}, 
February 2008, Baeza, Spain}

\vspace{1cm}

{\sc 

S.~Heinemeyer%
\footnote{
email: Sven.Heinemeyer@cern.ch}
}

\vspace*{1cm}

{\it
Instituto de Fisica de Cantabria (CSIC-UC), 
Santander,  Spain
}
\end{center}

\vspace*{0.2cm}

\BC {\bf Abstract} \EC
In these lecture notes 
we review some prospect for the upcoming LHC experiments in view of the
exploration of the Standard Model (SM) or its minimal Supersymmetric
extension (MSSM). We focus on some theoretical aspects concerning the
Higgs sector of the two models. We give results for the precision
observables $\MW$ and $\mt$ and their impact on the indirect
determination of the Higgs sector. We furthermore review some prospects for
the direct measurements in the SM and MSSM Higgs sector.

\def\thefootnote{\arabic{footnote}}
\setcounter{footnote}{0}

\newpage


\title{Higgs Physics at the LHC: Some Theory Aspects
}
\author{S. Heinemeyer
\address{IFCA (CSIC -- UC), 39005 Santander, Spain}
}
\maketitle
\begin{abstract}
In these lecture notes 
we review some prospect for the upcoming LHC experiments in view of the
exploration of the Standard Model (SM) or its minimal Supersymmetric
extension (MSSM). We focus on some theoretical aspects concerning the
Higgs sector of the two models. We give results for the precision
observables $\MW$ and $\mt$ and their impact on the indirect
determination of the Higgs sector. We furthermore review some prospects for
the direct measurements in the SM and MSSM Higgs sector.

\end{abstract}
  

\section{Introduction}
\label{sec:intro}

Identifying the mechanism of electroweak symmetry
breaking will be one of the main goals of the LHC. 
Many possibilities have been studied in the literature, of which 
the most popular ones are the Higgs mechanism within the Standard Model
(SM)~\cite{sm} and within the Minimal Supersymmetric Standard Model
(MSSM)~\cite{mssm}. 
Theories based on Supersymmetry (SUSY)~\cite{mssm} are widely
considered as the theoretically most appealing extension of the
SM. They are consistent with the approximate
unification of the gauge coupling constants at the GUT scale and
provide a way to cancel the quadratic divergences in the Higgs sector
hence stabilizing the huge hierarchy between the GUT and the Fermi
scales. Furthermore, in SUSY theories the breaking of the electroweak
symmetry is naturally induced at the Fermi scale, and the lightest
supersymmetric particle can be neutral, weakly interacting and
absolutely stable, providing therefore a natural solution for the dark
matter problem.
SUSY predicts the existence of scalar partners $\tilde{f}_L,
\tilde{f}_R$ to each SM chiral fermion, and spin--1/2 partners to the
gauge bosons and to the scalar Higgs bosons. 
The Higgs sector of the MSSM 
with two scalar doublets accommodates five physical Higgs bosons. In
lowest order these are the light and heavy $\cp$-even $h$
and $H$, the $\cp$-odd $A$, and the charged Higgs bosons $H^\pm$.

Other (non-SUSY) new physics models (NPM) that have been investigated in
the last decade comprise Two Higgs Doublet Models
(THDM) \cite{thdm}, little Higgs models~\cite{lhm}, or models with
(large, warped, \ldots) extra dimensions~\cite{edm}. 
However, we will restrict ourselves to the MSSM when discussing the LHC
capabilities for exploring physics beyond the SM.

So far, the direct search for NPM particles has not been successful.
One can only set lower bounds of ${\cal O}(100)$~GeV on
their masses~\cite{pdg}. The search reach is currently extended in various
ways in the ongoing Run~II at the Tevatron~\cite{TevFuture}. 
The increase in the reach for new physics phenomena, unfortunately, is
rather restricted. However, in autumn/winter of 2008 the
LHC~\cite{atlas,cms} is scheduled to start operation. With a center of
mass energy about seven times higher than the Tevatron it will
be able to test many realizations at the TeV scale (favored by naturalness
arguments) of the above mentioned ideas of NPM.
In the more far future, the $e^+e^-$ International Linear Collider
(ILC)~\cite{teslatdr,orangebook,acfarep} has very good prospects for
exploring the NPM in the per-cent range.
From the interplay of the LHC and the ILC detailed
information on many NPM can be expected in this case~\cite{lhcilc}.
Besides the direct detection of NPM particles (and Higgs bosons), 
physics beyond the SM can also be probed by precision observables via the
virtual effects of the additional particles.
This may permit to distinguish between e.g.\ the SM and the MSSM. 
However, this requires a very high precision
of the experimental results as well as of the theoretical predictions.

In theses lecture notes in \refse{sec:mwmt} we will briefly describe the
prospects of the 
measurements of the $W$~boson mass, $\MW$, and the top quark mass,
$\mt$, at the LHC and their implications for the SM and the MSSM. Theory
aspects of SM Higgs physics concerning mass and coupling constant
determination at the LHC will be presented in
\refse{sec:higgs}. In \refse{sec:hHA} we review some theory issues for
the searches for SUSY Higgs bosons at the LHC.


\section{\boldmath{$W$} boson and top quark mass measurements at the LHC}
\label{sec:mwmt}

As a first example for the LHC prospects we analyze the measurement of
the $W$~boson mass, $\MW$, and the top quark mass, $\mt$ and their
implications for the SM and the MSSM Higgs sector.

The top quark mass is a fundamental parameter of the SM (or the MSSM). 
So far it has been measured exclusively at the Tevatron,
yielding a precision of~\cite{mt1726,TEVEWWG}
\begin{align}
\label{mtexp}
\mt^{\rm exp} = 172.6 \pm 1.4 \gev~.
\end{align}
The corresponding Tevatron production cross sections have recently been
re-evaluated in \citere{sigmattunc}.
For $\MW$ progress has been achieved over the last
decade in the experimental measurements as well as in the theory
predictions in the SM and in the MSSM. The current experimental
value~\cite{LEPEWWG,TEVEWWG,MWcdf,gruenewald07} (see also \citere{ssdd})
\BE
\MW^{\rm exp} = 80.398 \pm 0.025 \gev
\EE
is based on a combination of the LEP results~\cite{LEP4f,lepewwg} and the
latest CDF measurement~\cite{MWcdf,gruenewald07}. 
The experimental measurement of $\MW$ also required substantial theory
input such as cross section evaluations for LEP~\cite{racoonww,yfsww} or
kinematics of $W$ and $Z$ boson decays~\cite{resbos} or the inclusion of
initial and final state photons~\cite{wgrad} at the Tevatron.
The current accuracies of $\mt$ and $\MW$ are summarized in
\refta{tab:POfuture}, together with their future expectations. 
Also included for completeness is the effective leptonic weak mixing
angle, for which hardly any improvement can be expected neither from the
Tevatron nor from the LHC.

\begin{table}[htb!]
\renewcommand{\arraystretch}{1.5}
\begin{center}

\begin{tabular}{|c||c|c|c|}
\cline{2-4} \multicolumn{1}{c||}{}
& now & Tevatron & LHC \\
\hline\hline
$\de\sweff(\times 10^5)$ & 16   & --- & 14--20 \\
\hline
$\de\MW$ [MeV]           & 25   &  20 & 15   \\
\hline
$\de\mt$ [GeV]           &  1.4 &  1.2 &  1.0 \\
\hline
$\de\MHSM/\MHSM$ [\%]     &  36 &     &  28 \\
\hline
\end{tabular}

\end{center}
\renewcommand{\arraystretch}{1}
\vspace{-1em}
\caption{
Current and anticipated future experimental uncertainties for
$\sweff$, $\MW$ and $\mt$. Also shown is the relative precision of the
indirect determination of $\MHSM$~\cite{gruenewald07}.
Each column represents the combined results of all detectors and
channels at a given collider, taking into account correlated
systematic uncertainties, see \citeres{blueband,gigaz,moenig,mwgigaz}
for details. 
}
\label{tab:POfuture}
\vspace{1em}
\end{table}

The importance of a precise $\mt$ and $\MW$ measurement comes from the
fact that $\MW$ can be calculated within the SM or the MSSM.
The theory prediction for the $W$~boson mass can be evaluated from
\BE
\MW^2 \KL 1 - \frac{\MW^2}{\MZ^2}\right) = 
\frac{\pi \al}{\sqrt{2} \GF} \left(1 + \De r\KR ,
\label{eq:delr}
\EE
where $\al$ is the fine structure constant and $\GF$ the Fermi constant.
The radiative corrections are summarized in the quantity 
$\De r$~\cite{sirlin}.
Within the SM the one-loop~\cite{sirlin} and the complete two-loop
result has been obtained for 
$\MW$~\cite{MWSM2L,MWSM}. 
Higher-order QCD
corrections are known at
\order{\al\als^2}~\cite{drSMgfals2,MWSMQCD3LII}. 
Leading electroweak contributions of order
\order{\gf^2 \als \mt^4} and \order{\gf^3 \mt^6} that enter via the
quantity $\De\rho$~\cite{rho} have been calculated in
\citeres{drSMgf3mh0,drSMgf3,drSMgf3MH}. 
The class of four-loop contributions obtained in 
\citere{deltarhoSM4L} give rise to a numerically negligible effect.
The prediction for $\MW$ within the SM
(or the MSSM) is obtained by evaluating $\De r$ in these models and
solving \refeq{eq:delr} for $\MW$. 

Within the MSSM the most precise available result for $\MW$ has been 
obtained in~\citere{MWpope}. Besides the full SM result, for the MSSM it
includes the full set of one-loop
contributions~\cite{deltarMSSM1lA,deltarMSSM1lB,MWpope}   
as well as the corrections of \order{\al\als}~\cite{dr2lA} and of
\order{\al_{t,b}^2}~\cite{drMSSMal2B,drMSSMal2} to the quantity
$\De\rho$, see~\citere{MWpope} for details.

\begin{figure}[htb!]
\begin{center}
\includegraphics[width=.80\textwidth,height=8cm]
                {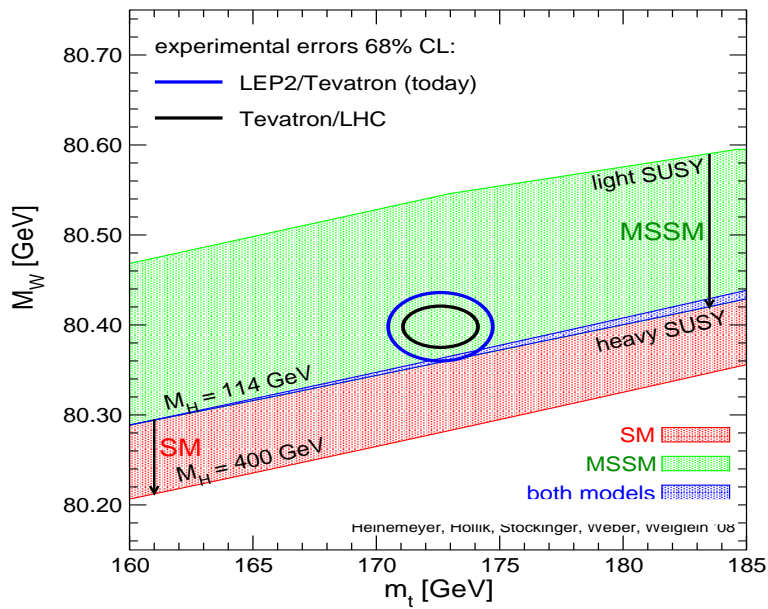}
\begin{picture}(0,0)
\CBox(-165,032)(-15,041){White}{White}
\end{picture}
\caption{%
Prediction for $\MW$ in the MSSM and the SM (see text) as a function of
$\mt$ in comparison with the present experimental results for $\MW$ and
$\mt$~\cite{MWpope}.
}
\label{fig:MWMTtoday}
\end{center}
\vspace{-2em}
\end{figure}

The experimental result and the theory prediction of the SM and the MSSM
are compared in \reffi{fig:MWMTtoday}.%
\footnote{
The plot shown here is an update of 
\citeres{deltarMSSM1lA,PomssmRep,MWpope}.
}%
~The predictions within the two models 
give rise to two bands in the $\mt$--$\MW$ plane with only a relatively small
overlap sliver (indicated by a dark-shaded (blue) area in
\reffi{fig:MWMTtoday}).  
The allowed parameter region in the SM (the medium-shaded (red)
and dark-shaded (blue) bands) arises from varying the only free parameter 
of the model, the mass of the SM Higgs boson, from $\MHSM = 114\gev$,
the LEP exclusion bound~\cite{LEPHiggsSM,LEPHiggsMSSM}
(upper edge of the dark-shaded (blue) area), to $400 \gev$ (lower edge
of the medium-shaded (red) area).
The light shaded (green) and the dark-shaded (blue) areas indicate 
allowed regions for the unconstrained MSSM, obtained from scattering the
relevant parameters independently~\cite{MWpope}. 
The decoupling limit with SUSY masses of \order{2 \tev}
yields the lower edge of the dark-shaded (blue) area. Thus, the overlap 
region between
the predictions of the two models corresponds in the SM to the region
where the Higgs boson is light, i.e.\ in the MSSM allowed region 
($\Mh \lsim 135 \gev$~\cite{mhiggslong,mhiggsAEC}, see
\refse{sec:h}). In the MSSM it 
corresponds to the case where all 
superpartners are heavy, i.e.\ the decoupling region of the MSSM.

\begin{figure}[htb!]
\vspace{-1em}
\begin{center}
\includegraphics[width=.80\textwidth,height=8cm]{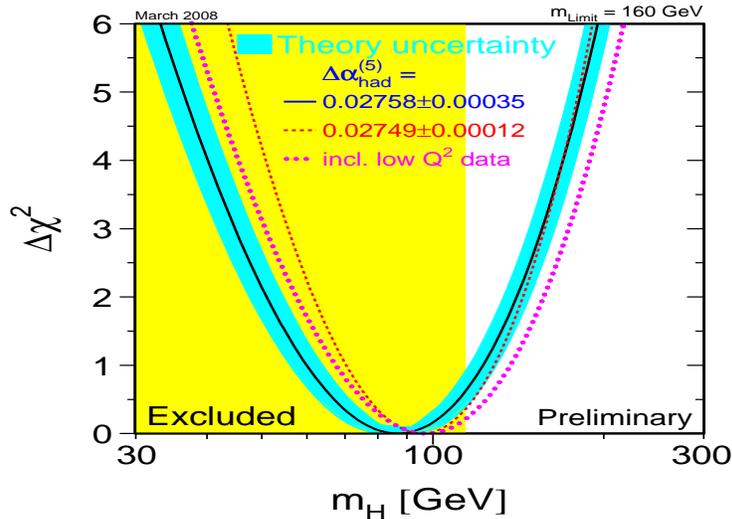}
\vspace{-2em}
\caption{%
$\De\chi^2$ curve derived from all EWPO measured at LEP, SLD, CDF and
D0, as a function of $\MHSM$, assuming the SM to be the correct theory
of nature~\cite{LEPEWWG}. 
}
\label{fig:blueband}
\end{center}
\vspace{-1em}
\end{figure}

The current 68\%~C.L.\ experimental results and their LHC expectations
for $\mt$ and $\MW$ are indicated in the plot as blue and black
ellipses, respectively. As can be seen from 
\reffi{fig:MWMTtoday}, the current experimental 68\%~C.L.\ region for 
$\mt$ and $\MW$ exhibits a slight preference of the MSSM over the SM.
At the 95\%~C.L.\ (not shown) the current experimental values enter the
SM parameter 
space. This example indicates that the experimental measurement of $\MW$
in combination with $\mt$ 
prefers within the SM a relatively small value of $\MHSM$, or with the
MSSM not too heavy SUSY mass scales.
A fit to all electroweak precision observables (EWPO) (including $\MW$
and $\mt$) determines $\MHSM$
currently to $\sim 36\%$~\cite{LEPEWWG}, 
\begin{align}
\label{MHSMfit}
\MHSM = 87^{+36}_{-27} \gev .
\end{align}
This is shown in the ``blue band'' plot in
\reffi{fig:blueband}~\cite{LEPEWWG}. In this figure $\De\chi^2$ is shown as a
function of $\MHSM$, yielding \refeq{MHSMfit} as best fit with an upper
limit of $160 \gev$ at 95\%~C.L. This value increases to $190 \gev$ if
the direct LEP bound of $114.4 \gev$ at the 95\%~C.L.~\cite{LEPHiggsSM}
is included in the fit. The 
theory (intrinsic) uncertainty in the SM calculations (as evaluated with 
{\tt TOPAZ0}~\cite{topaz0} and {\tt ZFITTER}~\cite{zfitter}) are
represented by the thickness of the blue band. The width of the parabola
itself, on the other hand, is determined by the experimental precision of
the measurements of the EWPO and the input parameters.

With the future improvements at the LHC in $\mt$ and $\MW$, shown as the
black ellipse in \reffi{fig:MWMTtoday}, the precision
of the indirect determination of $\MHSM$ will
increase~\cite{gruenewald07} as indicated in \refta{tab:POfuture}.


\section{SM Higgs physics at the LHC}
\label{sec:higgs}

\subsection{Discovering a Higgs boson}

In order to ``discover the Higgs boson'' several steps have to be
taken, as summarized in \refta{tab:higgsdiscovery}. Simply detecting a
new particle and measure its mass (and checking whether it is in
agreement with the model predictions, see e.g.\ the last line of
\refta{tab:POfuture}) is not sufficient. In order to
establish the Higgs mechanism the coupling of the new state to fermions
and gauge bosons has to be measured and compared with the model
predictions. Finally also the self-coupling (corresponding to the Higgs
potential) as well as its spin and quantum numbers have to be determined
experimentally. Only if all measurements agree with the prediction of
the Higgs mechanism (of e.g.\ the SM) ``discovery of the Higgs boson''
can be claimed. 

Finding a Higgs candidate particle and providing a rough mass
measurement might still be possible at the Tevatron (depending somewhat
on its mass)~\cite{TevFuture}. The discovery of a SM-like Higgs is
guaranteed at the LHC, where also a relative precise mass measurement
and possibly a coarse measurement of its couplings to SM fermions and
gauge bosons can be performed, see \refse{sec:SMhiggs}. However, the
final steps for the establishment of the Higgs mechanism, measurement of
the self-coupling as well as of the quantum numbers can most probably
only be performed at the $e^+ e^-$
ILC~\cite{teslatdr,orangebook,acfarep,Snowmass05Higgs}. The anticipated
capabilities of the various colliders are summarized in
\refta{tab:higgsdiscovery}. 

\begin{table}[htb!]
\renewcommand{\arraystretch}{1.5}
\begin{center}
\begin{tabular}{lccc}
1. Find the new particle 
& T & L & I \\
2. measure its mass ($\Rightarrow$ ok?)
& T & L & I \\
3. measure coupling to gauge bosons
&  & L & I \\
4. measure couplings to fermions
&  & L & I \\
5. measure self-couplings
&  &   & I \\
6. measure spin, \ldots
&  &   & I \\
\end{tabular}
\end{center}
\renewcommand{\arraystretch}{1}
\caption{
Steps that have to be taken to ``discover the Higgs''. Indicated are the
colliders at which a certain measurement (most likely) can be performed: 
T = Tevatron, L = LHC, I = ILC.
}
\label{tab:higgsdiscovery}
\end{table}


\subsection{SM Higgs boson mass and couplings at the LHC}
\label{sec:SMhiggs}

A SM-like Higgs boson can be produced in many channels at the LHC as
shown in \reffi{fig:LHCHiggsXS} (taken from \citere{sigmaH}, where also
the relevant original references can be found).
The corresponding discovery potential for a SM-like Higgs boson of
ATLAS is shown in \reffi{fig:ATLASHiggsdiscovery}~\cite{AtlasHiggs},
where similar results have been obtained for CMS~\cite{cms}.
With $10\,\ifb$ a $5\,\si$ discovery is expected for 
$\MHSM \gsim 130 \gev$. For lower masses a higher integrated luminosity
will be needed. 
The largest production cross section is reached by $gg \to H$, which
however, will be visible only in the decay to SM gauge bosons. A precise
mass measurement of $\Mh^{\rm exp} \approx 200 \mev$ can be provided by
the decays $H \to \ga\ga$ at lower 
Higgs masses and by $H \to ZZ^{(*)} \to 4\ell$ at higher masses. 
This guarantees the detection of the new state and a precise mass measurement
over the relevant parameter space within the SM.

\medskip
The next step will be the determination of the Higgs boson couplings to
SM fermions and gauge bosons. We will focus here on the analysis
presented in \citere{HcoupSM}.
As shown in \reffi{fig:LHCHiggsXS}, the LHC will
provide us with many different Higgs observation channels.  In the SM
there are four relevant production modes: gluon fusion (GF;
loop-mediated, dominated by the top quark), which dominates
inclusive production; weak boson fusion (WBF), which has an
additional pair of hard and far-forward/backward jets in the final
state; top-quark associated production ($t\bar{t}H$); and weak boson
associated production ($WH,ZH$), where the weak boson is identified by
its leptonic decay.
In general, the LHC
will be able to observe Higgs decays to photons, weak bosons, tau
leptons and $b$ quarks, in the range of Higgs masses where the
branching ratio (BR) in question is not too small.

\begin{figure}[htb!]
\vspace{-1em}
\begin{center}
\includegraphics[angle=-90,width=11cm]{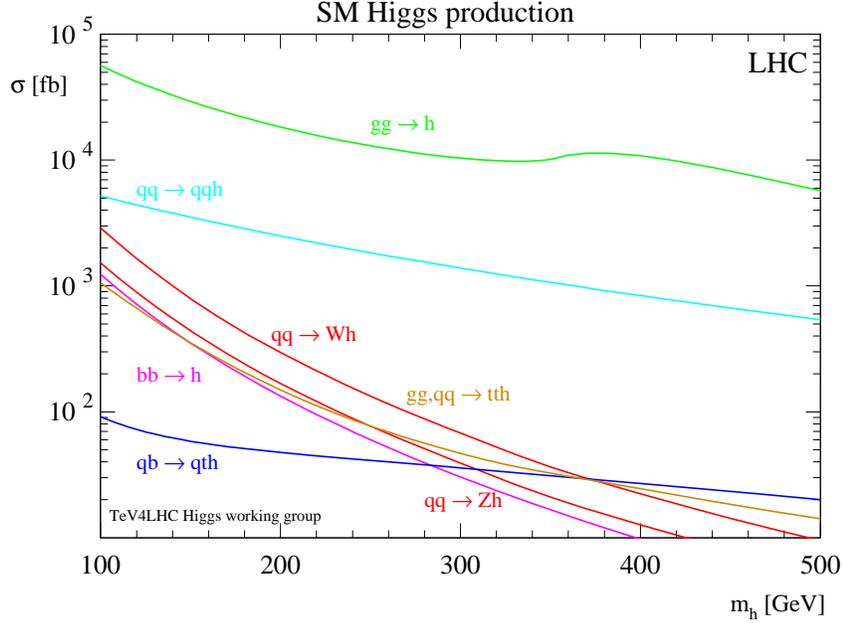}
\caption{%
The various production cross sections for a SM-like Higgs boson at the
LHC are shown as a function of $\MHSM$ (taken from \citere{sigmaH},
where also the relevant references can be found).
}
\label{fig:LHCHiggsXS}
\end{center}
\end{figure}
%
\begin{figure}[b!]
\vspace{-2em}
\begin{center}
\includegraphics[width=12cm]{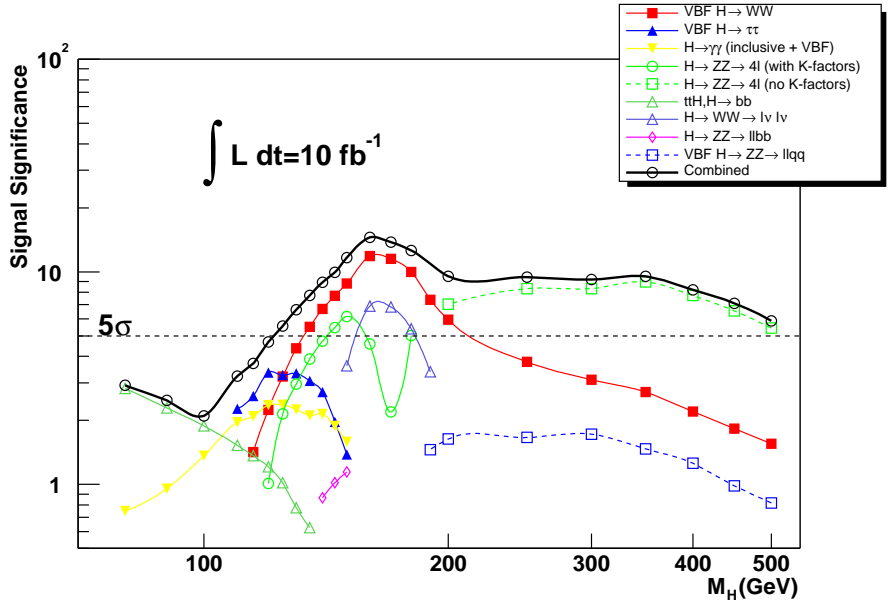}
\caption{%
Significance of a Higgs signal, measured at ATLAS with 
$10\,\ifb$~\cite{AtlasHiggs}.
Similar results have been obtained for CMS~\cite{cms}.
}
\label{fig:ATLASHiggsdiscovery}
\end{center}
\vspace{-3em}
\end{figure}

For a Higgs in the intermediate mass range, the total width, $\Gamma$,
is expected to be small enough to use the narrow-width approximation
in extracting couplings.  The rate of any channel (with the $H$
decaying to final state particles $xx$) is, to a good approximation,
given by
\begin{equation}
\sigma(H) \times {\rm BR} (H\to xx) = 
{\sigma(H)^{\rm SM}\over \Gamma_p^{\rm SM}}
\cdot {\Gamma_p\Gamma_x \over \Gamma}\;,
\end{equation}
where $\Gamma_p$ is the Higgs partial width involving the production
couplings, and where the Higgs branching ratio for the decay is written
as ${\rm BR}(H\to xx)=\Gamma_x/\Gamma$. Even with cuts, the observed
rate directly determines the product $\Gamma_p\Gamma_x/\Gamma$
(normalized to the calculable SM value of this product).  The LHC will
have access to (or provide upper limits on) combinations of
$\Gamma_g,\Gamma_W,\Gamma_Z, \Gamma_\gamma,\Gamma_\tau,\Gamma_b$ and
the square of the top Yukawa coupling, $Y_t$.
The analysis of \citere{HcoupSM} was based on the channels:
GF $gg \rightarrow H \rightarrow Z Z$,
WBF $qq \, H \rightarrow qq\, Z Z$,
GF $gg \rightarrow H \rightarrow W W$,
WBF $qq \, H \rightarrow qq\, W W$,
$W \, H \rightarrow W\, W W$ (2$l$ and 3$l$ final state),
$t\bar{t} \, H (H \rightarrow W W, t \rightarrow W b)$ (2$l$ and 3$l$
final state),
inclusive Higgs boson production: $H \rightarrow \gamma \gamma$,
WBF $qq \, H \rightarrow qq\, \gamma \gamma$,
$t\bar{t} \, H (H \rightarrow \gamma \gamma)$,
$W \, H (H \rightarrow \gamma \gamma)$,
$Z \, H (H \rightarrow \gamma \gamma)$,
WBF $qq \, H \rightarrow qq\, \tau \tau$,
$t\bar{t} \, H (H \rightarrow b\bar{b})$.
The significance of the last channel has
become substantially worse in the recent ATLAS and CMS
analyses~\cite{AtlasHiggs,cms} (also the other channels have been
re-analyzed in the recent years). This should be kept in mind for the 
results reviewed here. 
They might become worse in an updated analysis.

The production and decay channels listed above refer to a single
Higgs resonance, with decay signatures which also exist in the SM. The
Higgs sector may be much richer, of course, see e.g.\ \refse{sec:hHA}.  
However, no analysis has
yet been performed for a non-SM-like Higgs scenario.

While from the channels listed above ratios of couplings (or partial
widths) can be extracted in a fairly model-independent way, see e.g.\ 
\citere{ATL-PHYS-2003-030}, further theoretical
assumptions are 
necessary in order to determine absolute values of the Higgs couplings
to fermions and bosons and of the total Higgs boson width%
\footnote{
An assumption free determination of Higgs boson couplings will be
possible at the ILC.}%
. The only
assumption that was used in \citere{HcoupSM} is that the strength of
the Higgs--gauge-boson couplings does not exceed the SM value by more
than 5\%
\begin{equation}
\Gamma_V\leq\Gamma_V^{\rm SM} \times 1.05, \quad V=W,Z~.
\label{eq:constraint}
\end{equation}
This assumption is justified in any model with an arbitrary number of
Higgs doublets
(with or without additional Higgs singlets), i.e., it is true for the
MSSM in particular.
While \refeq{eq:constraint} constitutes an upper bound on the Higgs
coupling to weak bosons, the mere observation of Higgs production
puts a lower bound on the production couplings and, thereby, on the
total Higgs width. The constraint $\Gamma_V\leq\Gamma_V^{\rm SM} \times 1.05$,
combined with a measurement of $\Gamma_V^2/\Gamma$ from observation of
$H\to VV$ in WBF, then puts an upper bound on the Higgs total width,
$\Gamma$. Thus, an absolute determination of the Higgs total width is
possible in this way. Using this result, an absolute determination also
becomes possible for Higgs couplings to gauge bosons and
fermions.

The expected LHC accuracies are obtained from a $\chi^2$~fit based on 
experimental information for the channels listed above. 
Details on the fitting procedure, error assumptions etc.\ can be found
in \citere{HcoupSM}.
The results below are shown for two luminosity assumptions for the
LHC:\\[.3em]
      30 fb$^{-1}$ at each of two experiments, denoted 
\underline{$2 \times 30$~fb$^{-1}$};\\[.3em]
      300 fb$^{-1}$ at each of two experiments, of which only 100 fb$^{-1}$
is usable for WBF channels at each experiment, denoted 
\underline{$2 \times 300$ $+$ $2 \times 100$~fb$^{-1}$};\\[.3em]
The latter case allows for possible significant degradation of the WBF
channels in a high luminosity environment.

\begin{figure}[htb!]
\begin{center}
\includegraphics[width=6cm]{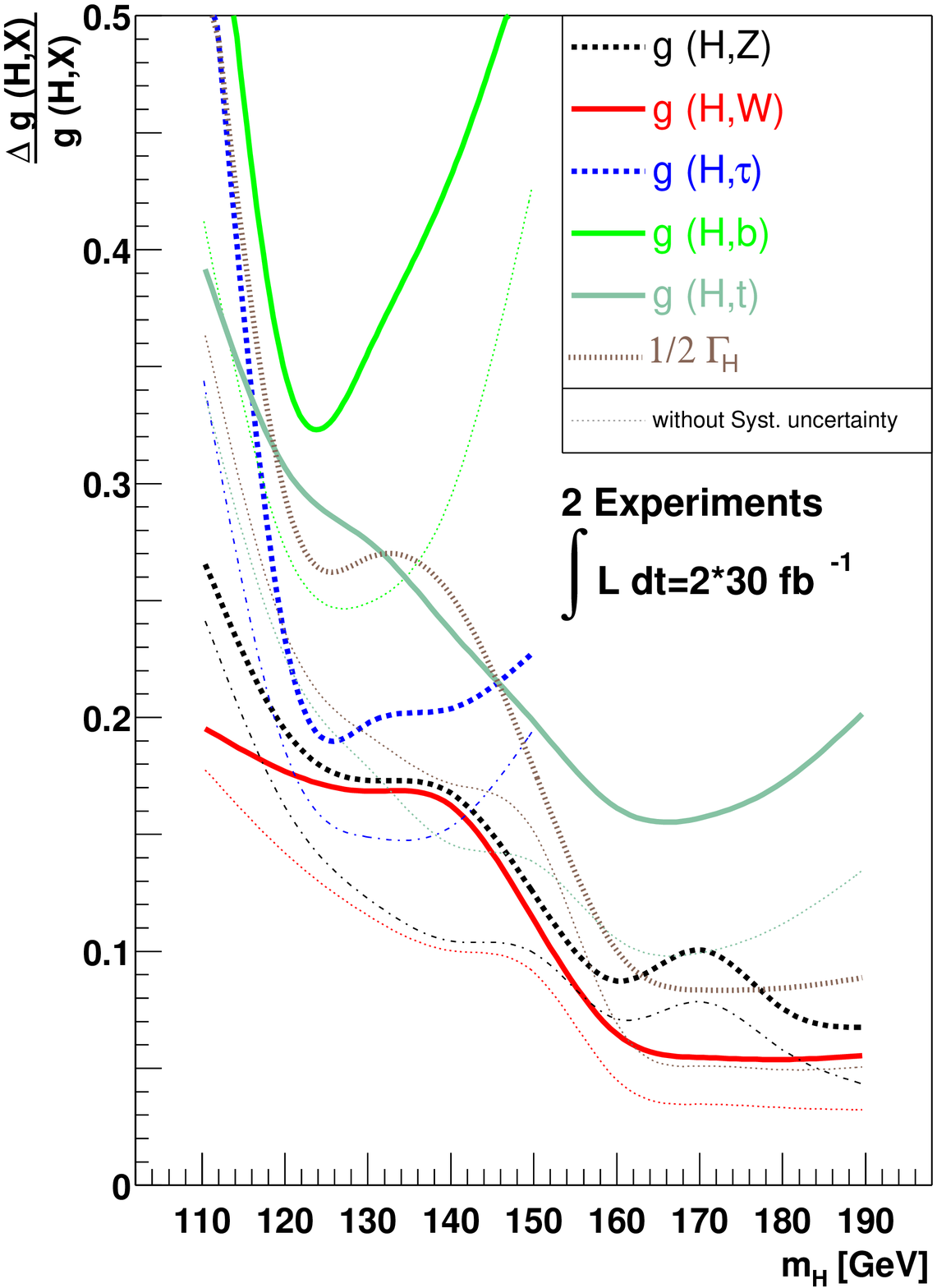}
\includegraphics[width=6cm]{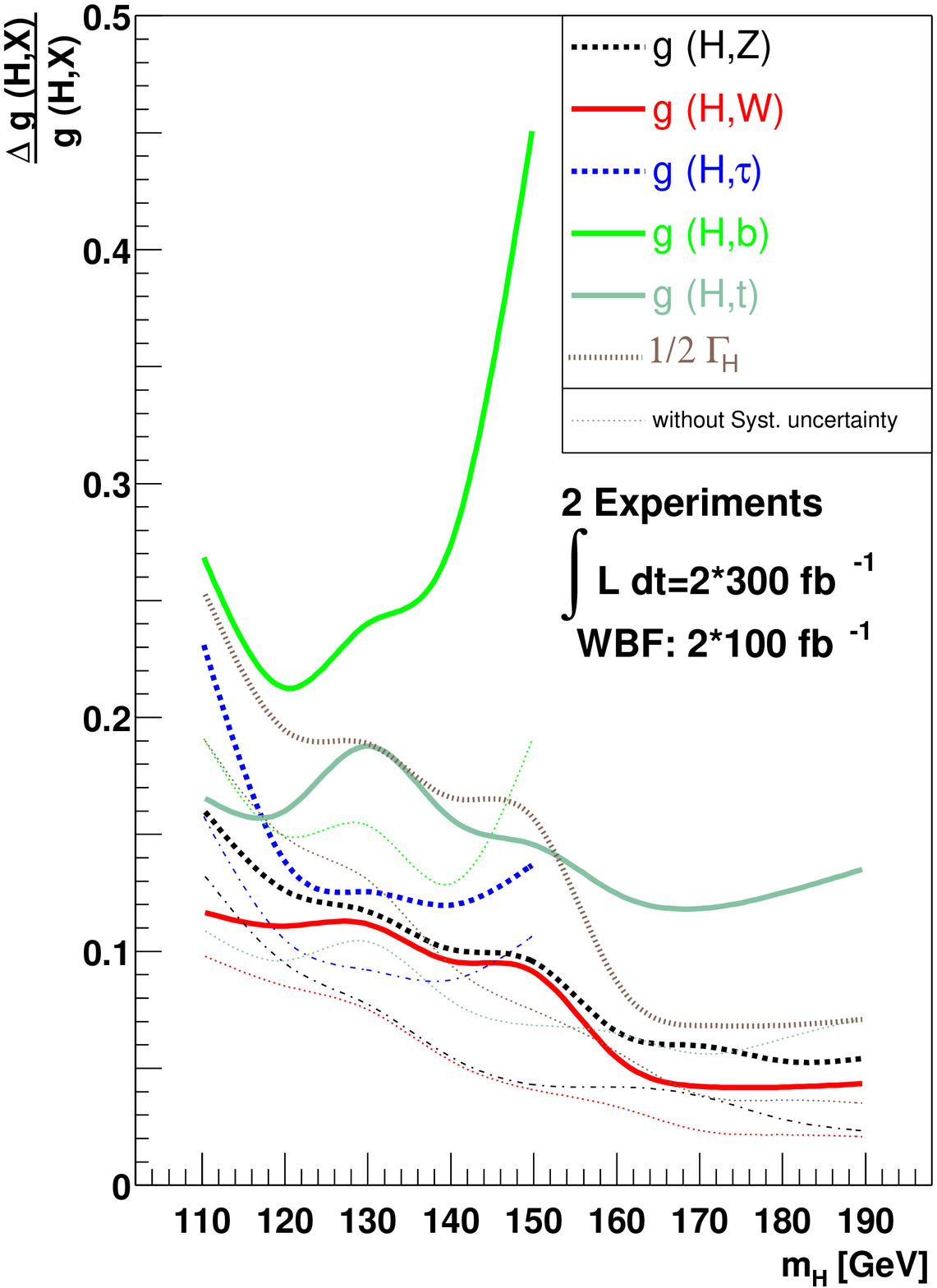}
\caption{%
Precision for various Higgs boson couplings based on \citere{HcoupSM} as
a function of the Higgs boson mass.
The left plot assumes $30\,\ifb$ at each of the two experiments, the
right plot assumes $300\,\ifb$ at each experiment, where, however, only 
$100\,\ifb$ can be used in the WBF channel. 
The thin lines show the result without systematic uncertainties.
}
\label{fig:HcoupSM}
\end{center}
\vspace{-1em}
\end{figure}

The results of the fit for the Higgs boson couplings are shown in
\reffi{fig:HcoupSM} as a function of the SM Higgs boson mass.
The left plot contains the results for the low luminosity
case. Couplings to gauge bosons can be determined at the level of 
$\sim 20\%$ for $\MHSM \lsim 150 \gev$ and at 5--10\% for higher Higgs
boson masses. For the couplings to SM fermions the fit yields a
precision between 20~and 40\% depending on the fermion species and
$\MHSM$. Above $\MHSM \approx 150 \gev$ the only fermion coupling that 
can be determined is the one to the top quark.
The accuracies of the coupling determination improve substantially for
the high luminosity case as shown in the right plot of
\reffi{fig:HcoupSM}. Couplings to the SM gauge bosons are fitted with a
precision of 5--10\% for all $\MHSM$, and the couplings to fermions are
measured a the 15--25\% level.

As mentioned above, taking recent experimental analyses on the various
Higgs production and decay channels into account~\cite{AtlasHiggs,cms},
the results could change. 
Furthermore there might be some chances to measure the spin, the $\cp$
properties and the $HVV$ vertex structure of the Higgs bosons for 
$\MHSM \gsim 160 \gev$~\cite{LHCfurtherHiggs}. However, within the SM
this mass range is disfavored by the electroweak precision data, see
\refse{sec:mwmt}.


\section{MSSM Higgs bosons at the LHC}
\label{sec:hHA}

In this section we focus on the MSSM with real parameters.
Contrary to the Standard Model (SM), in the MSSM two Higgs doublets
are required.
The  Higgs potential
\BEA
V &=& m_{1}^2 |\cHe|^2 + m_{2}^2 |\cHz|^2 
      - m_{12}^2 (\epsilon_{ab} \cHe^a\cHz^b + \mbox{h.c.})  \non \\
  & & + \frac{1}{8}(g_1^2+g_2^2) \left[ |\cHe|^2 - |\cHz|^2 \right]^2
        + \frac{1}{2} g_2^2|\cHe^{\dag} \cHz|^2~,
\label{higgspot}
\EEA
contains $m_1, m_2, m_{12}$ as soft SUSY breaking parameters;
$g_2, g_1$ are the $SU(2)$ and $U(1)$ gauge couplings, and 
$\epsilon_{12} = -1$.

The doublet fields $H_1$ and $H_2$ are decomposed  in the following way:
\BEA
\cHe &=& \VL \cHe^0 \\[0.5ex] \cHe^- \VR \; = \; \VL v_1 
        + \frac{1}{\sqrt2}(\phi_1^0 - i\chi_1^0) \\[0.5ex] -\phi_1^- \VR~,  
        \non \\
\cHz &=& \VL \cHz^+ \\[0.5ex] \cHz^0 \VR \; = \; \VL \phi_2^+ \\[0.5ex] 
        v_2 + \frac{1}{\sqrt2}(\phi_2^0 + i\chi_2^0) \VR~.
\label{higgsfeldunrot}
\EEA
The potential (\ref{higgspot}) can be described with the help of two  
independent parameters (besides $g_2$ and $g_1$): 
$\Tb = v_2/v_1$ and $M_A^2 = -m_{12}^2(\Tb+\CTb)$,
where $M_A$ is the mass of the $\cp$-odd Higgs boson~$A$.

The diagonalization of the bilinear part of the Higgs potential,
i.e.\ of the Higgs mass matrices, is performed via the orthogonal
transformations 
\BEA
\label{hHdiag}
\VL H^0 \\[0.5ex] h^0 \VR &=& \ML \Ca & \Sa \\[0.5ex] -\Sa & \Ca \MR 
\VL \phi_1^0 \\[0.5ex] \phi_2^0~, \VR  \\
\label{AGdiag}
\VL G^0 \\[0.5ex] A^0 \VR &=& \ML \Cb & \Sbe \\[0.5ex] -\Sbe & \Cb \MR 
\VL \chi_1^0 \\[0.5ex] \chi_2^0 \VR~,  \\
\label{Hpmdiag}
\VL G^{\pm} \\[0.5ex] H^{\pm} \VR &=& \ML \Cb & \Sbe \\[0.5ex] -\Sbe & 
\Cb \MR \VL \phi_1^{\pm} \\[0.5ex] \phi_2^{\pm} \VR~.
\EEA
The mixing angle $\al$ is determined through
\BE
\al = {\rm arctan}\KKL 
  \frac{-(\MA^2 + \MZ^2) \Sbe \Cb}
       {\MZ^2 \CQb + \MA^2 \SQb - m^2_{h,{\rm tree}}} \KKR~, ~~
 -\frac{\pi}{2} < \al < 0~.
\label{alphaborn}
\end{equation}

One gets the following Higgs spectrum:
\BEA
\mbox{2 neutral bosons},\, {\cal CP} = +1 &:& h, H \non \\
\mbox{1 neutral boson},\, {\cal CP} = -1  &:& A \non \\
\mbox{2 charged bosons}                   &:& H^+, H^- \non \\
\mbox{3 unphysical Goldstone bosons}      &:& G, G^+, G^- .
\EEA

Since the MSSM Higgs boson sector at tree-level can be described with
$\MA$ and $\tb$, all other masses and mixing angles are predicted. These
tree-level relations receive large higher-order corrections, 
see e.g.\ \citeres{PomssmRep,habilSH,mhiggsAWB} for reviews.
A typical mass spectrum (including higher-order corrections, 
obtained with the Fortran code 
{\tt FeynHiggs}~\cite{mhiggslong,mhiggsAEC,mhcMSSMlong,feynhiggs}) is
shown in \reffi{fig:MSSM_MH}. The Higgs masses are shown as a function
of $\MA$ for $\tb$ in the $\mhmax$ benchmark
scenario~\cite{benchmark2}. At low $\MA$ the lightest Higgs boson mass
rises, until at around $\MA \approx 200 \gev$ a maximum and plateau is
reached.  
This is the so-called ``decoupling'' limit. Here the lightest Higgs is
SM-like, while all the masses of the heavy Higgses are very close to
each other.

\begin{figure}[htb!]
\vspace{1em}
\begin{center}
\includegraphics[width=11cm,height=8cm]{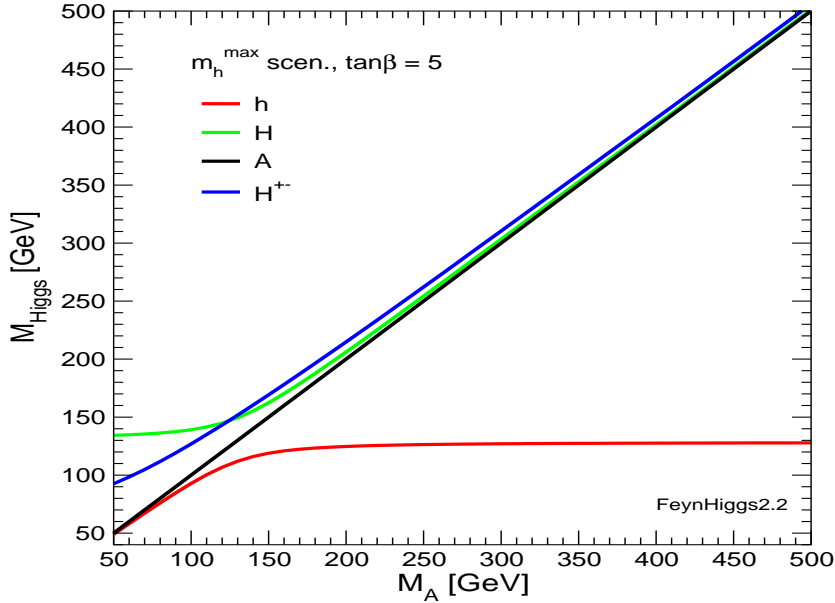}
\caption{%
The MSSM Higgs boson masses (including higher-order corrections) 
are shown as a function of $\MA$ for $\tb = 5$
in the $\mhmax$ benchmark scenario~\cite{benchmark2}.
}
\label{fig:MSSM_MH}
\end{center}
\end{figure}

The couplings of the MSSM Higgs bosons differ already at the tree-level
from the corresponding 
SM couplings. Some couplings important for the Higgs boson phenomenology
are given by
\begin{align}
g_{hVV} &= \sin(\be - \al) \; g_{HVV}^{\rm SM}, \quad V = W^{\pm}, Z , \\
g_{HVV} &= \cos(\be - \al) \; g_{HVV}^{\rm SM} , \\
\label{hbb}
g_{h b\bar b}, g_{h \tau^+\tau^-} &= - \sin\al/\cos\be \; 
   g_{H b\bar b, H \tau^+\tau^-}^{\rm SM} , \\
g_{h t\bar t} &= \cos\al/\sin\be \; g_{H t\bar t}^{\rm SM} , \\
\label{Abb}
g_{A b\bar b}, g_{A \tau^+\tau^-} &= \gamma_5\tan\be \; 
                                     g_{H b\bar b}^{\rm SM} .
\end{align}
The couplings of the neutral $\cp$-even Higgs bosons to gauge bosons is
always suppressed. However, not all of the couplings can be made small
simultaneously. The coupling of the light Higgs boson to down-type
fermions (especially to bottom quarks and tau leptons) can be enhanced
at large $\tb$. In the decoupling limit one finds in $\be - \al \to \pi/2$
and specifically $g_{hxx} \to g_{Hxx}^{\SM}$. On the other hand, 
the coupling of the $\cp$-odd Higgs boson to down-type fermions is always
enhanced by $\tb$. 

The resulting LHC production cross sections are shown in
\reffi{fig:LHC_MSSM_XS}~\cite{sigmaH}. They are given as a function of
$\MA$ for $\tb = 5$ in the $\mhmax$ benchmark scenario. The most
striking difference in comparison with the SM is the appearance of the 
$b\bar b\phi$ ($\phi = h, H, A$) channel. Due to the possible
enhancement in the MSSM, see \refeqs{hbb}, (\ref{Abb}), this cross
section can yield a detectable signal. 

\begin{figure}[htb!]
\begin{center}
\includegraphics[width=11cm,height=9cm]{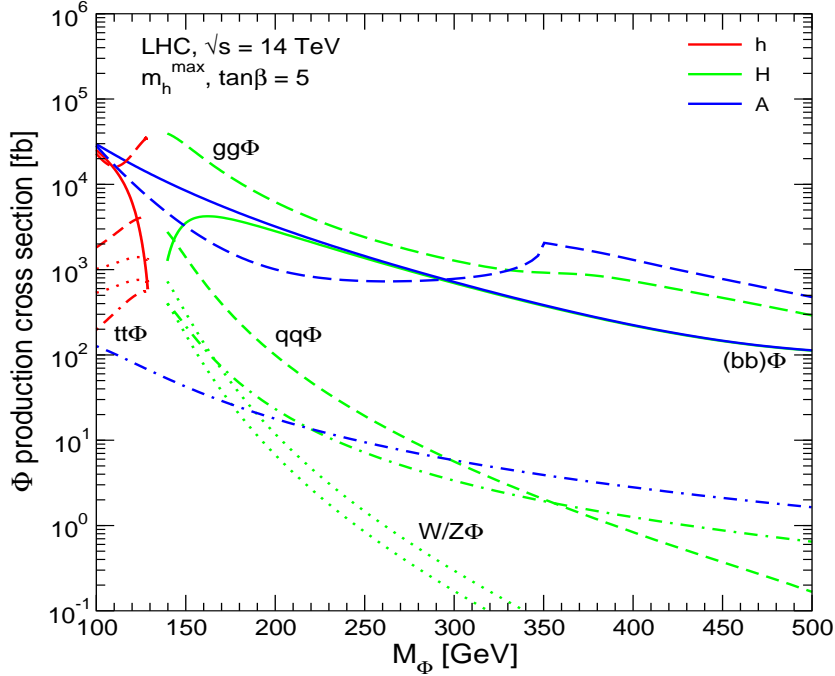}
\caption{%
Overview about the various neutral Higgs boson production cross sections
at the LHC shown as a function of $\MA$ for $\tb = 5$ in the $\mhmax$
scenario (taken from \citere{sigmaH}, where the original references can
be found).
}
\label{fig:LHC_MSSM_XS}
\end{center}
\vspace{-1em}
\end{figure}


\subsection{The light MSSM Higgs boson}
\label{sec:h}

The mass of the lightest MSSM Higgs boson mass is bounded from above by
$\MZ$ at the tree-level. Including loop corrections this bound is pushed
upwards to $\Mh \lsim 135 \gev$ (as obtained with 
{\tt FeynHiggs}). This
bound includes already the parametric uncertainty from the top-quark
mass, see \refeq{mtexp}, and the theory uncertainty due to unknown
higher-order corrections~\cite{mhiggsAEC,PomssmRep}. This bound makes a
firm prediction for the 
searches at the LHC.

Alternatively, $\Mh$ can be determined in a global fit to the EWPO,
similar to the result shown in \reffi{fig:blueband}. 
The corresponding result for $\Mh$ in the 
Constrained MSSM (CMSSM)%
\footnote{
The CMSSM is described by the parameters $m_{1/2}$,
$m_0$ and $A_0$ at the grand unification scale as well as $\tb$ and the
sign of the Higgs mixing parameter, sign$(\mu)$.
}%
~has been obtained in \citere{redband}. Furthermore included in
the fit are the anomalous magnetic moment of the muon, the B-physics
observables $\br(b \to s \ga)$ and $\br(B_s \to \mu^+\mu^-)$ as well as
the prediction for the Cold Dark Matter abundance, see \citere{redband}
for all the relevant references. This yields (using 
$\mt^{\rm exp} = 170.9 \pm 1.8 \gev$)
\begin{align}
\label{Mhfit}
\Mh^{\rm CMSSM} = 110^{+10}_{-8}\pm 3 \gev ,
\end{align}
where the first error is experimental and the second one due to unknown
higher-order corrections~\cite{mhiggsAEC,PomssmRep}. This has to be
compared to \refeq{MHSMfit} and 
to the bounds on the Higgs boson masses obtained at LEP, 
$114.4 \gev$ at the 95\%~C.L.~\cite{LEPHiggsSM}, which is valid also in
the CMSSM~\cite{asbs1,ehow1}. Despite its simplicity, the $\Mh$
prediction in the CMSSM is in better agreement with the LEP bounds than
the SM. 

Within the decoupling limit, $\MA \gsim 200 \gev$, the lightest MSSM
Higgs is SM-like. Its production and decay and consequently its
detection can proceed via the SM channels. However, two possible
deviations from the SM case should be kept in mind.
First, the $gg \to h$ cross section can be strongly suppressed due to
additional scalar top loops~\cite{gghsuppression}. This is realized in
the ``gluophobic Higgs'' benchmark scenario~\cite{benchmark2}. In this
case for  $\MA \lsim 400 \gev$ a suppression of the $gg \to h$
production cross section by 
more than 60\% as compared to the SM cross section takes place.
Second, the decay $h \to \ga\ga$ can be suppressed due to an enhanced 
$hb\bar b$ coupling. This could hamper the precise measurement of $\Mh$,
which proceeds via the decay to photons for $\Mh \lsim 130 \gev$. As an
example, with $30 \, \ifb$ in the $\mhmax$ scenario a precise mass
measurement will only be possible for $\MA \gsim 300 \gev$~\cite{cms}.

  
\subsection{The heavy MSSM Higgs bosons}
\label{sec:HA}

The main channel to discover the heavy neutral Higgs bosons for \\
$\MA \gsim 200 \gev$ is the production in association with bottom quarks
and the subsequent decay to tau leptons, 
$b \bar b \to b \bar b \; H/A \to b \bar b \; \tau^+\tau^-$.
For heavy supersymmetric particles, with masses far above
the Higgs boson mass scale, one has for the production and decay of the
$A$~boson~\cite{benchmark3} 
\begin{align}
\label{eq:bbA}
& \sigma(b\bar{b} A) \times {\rm BR}(A \to b \bar{b}) \simeq
\sigma(b\bar{b} H)_{\rm SM} \;
\frac{\tan^2\be}{\left(1 + \db \right)^2} \times
\frac{ 9}{
\left(1 + \db \right)^2 + 9} ~, \\
\label{eq:Atautau}
& \sigma(gg, b\bar{b} \to A) \times {\rm BR}(A \to \tau^+ \tau^-) \simeq
\sigma(gg, b\bar{b} \to H)_{\rm SM} \;
\frac{\tan^2\be}{
\left(1 + \db \right)^2 + 9} ~,
\end{align} 
where $\sigma(b\bar{b}H)_{\rm SM}$ and $\sigma(gg, b\bar{b} \to H)_{\rm SM}$ 
denote the values of the corresponding SM Higgs boson production cross
sections for $\MHSM = \MA$.
$\db$ is given by~\cite{deltamb1}
\BE
\db = \frac{2\als}{3\,\pi} \, \mgl \, \mu \, \tb \,
                    \times \, I(\msbe, \msbz, \mgl) +
      \frac{\alt}{4\,\pi} \, \At \, \mu \, \tb \,
                    \times \, I(\mste, \mstz, \mu) ~,
\label{def:dmb}
\end{equation}
where the function $I$ arises from the one-loop vertex diagrams and
scales as
$I(a, b, c) \sim 1/\mbox{max}(a^2, b^2, c^2)$.
Here $\mste, \mstz$ and $\msbe, \msbz$ denote the two scalar top and
bottom masses, respectively. $\mgl$ is the gluino mass, $\mu$ is the
Higgs mixing parameter, and $\At$ denotes the trilinear Higgs-stop
coupling. 
As a consequence, the $b\bar{b}$ production rate depends sensitively on
$\db \propto \mu\,\tb$ because of the factor $1/(1 + \db)^2$, while this
leading 
dependence on $\db$ cancels out in the $\tau^+\tau^-$ production rate.
The formulas above apply, within a good approximation, also to the
heavy $\cp$-even Higgs boson in the large $\tb$ regime. 
Therefore, the production and decay
rates of $H$ are governed by similar formulas as the ones given
above, leading to an approximate enhancement by a factor 2 of the production
rates with respect to the ones that would be obtained in the case of the
single production of the $\cp$-odd Higgs boson as given in
\refeqs{eq:bbA}, (\ref{eq:Atautau}). 

Of particular interest is the ``LHC wedge'' region, i.e.\
the region in which only the light $\cp$-even MSSM Higgs boson, but non
of the heavy MSSM Higgs bosons can be
detected at the LHC at the 5$\,\si$ level. It appears for 
$\MA \gsim 200 \gev$ at intermediate $\tb$ and widens to larger $\tb$
values for larger $\MA$. Consequently, in the ``LHC wedge'' only a
SM-like light Higgs 
boson can be discovered at the LHC. This region is bounded from above by
the $5\,\si$ discovery contours for the heavy neutral MSSM Higgs bosons
as described above. These discovery contours depend sensitively 
on the Higgs mass parameter $\mu$. The dependence on
$\mu$ enters in two different ways, on the one hand via higher-order
corrections through $\db \propto \mu\,\tb$, 
and on the other hand
via the kinematics of Higgs decays into charginos and neutralinos, where
$\mu$ enters in their respective mass matrices~\cite{mssm}. 

In \reffi{fig:LHCwedge} we show the $5\,\si$ discovery regions for the
heavy neutral MSSM Higgs bosons in the channel
$b \bar b \to b \bar b \; H/A , H/A \to \tau^+\tau^- \to \mbox{jets}$
\cite{cmsHiggs}. 
As explained above, these discovery contours correspond to the upper
bound of the ``LHC wedge''. A strong variation with the sign and the
size of $\mu$ can be observed and should be taken into account in
experimental and phenomenological analyses. 
The same higher-order corrections are relevant once a possible heavy
Higgs boson signal at the LHC will be interpreted in terms of the underlying
parameter space. From \refeq{def:dmb} it follows that an observed
production cross section can be correctly connected to $\mu$ and $\tb$
only if the scalar top and bottom masses, the gluino mass and the
trilinear Higgs-stop coupling are measured and taken properly into account.

\begin{figure}[htb!]
\begin{center}
\includegraphics[width=.45\textwidth]{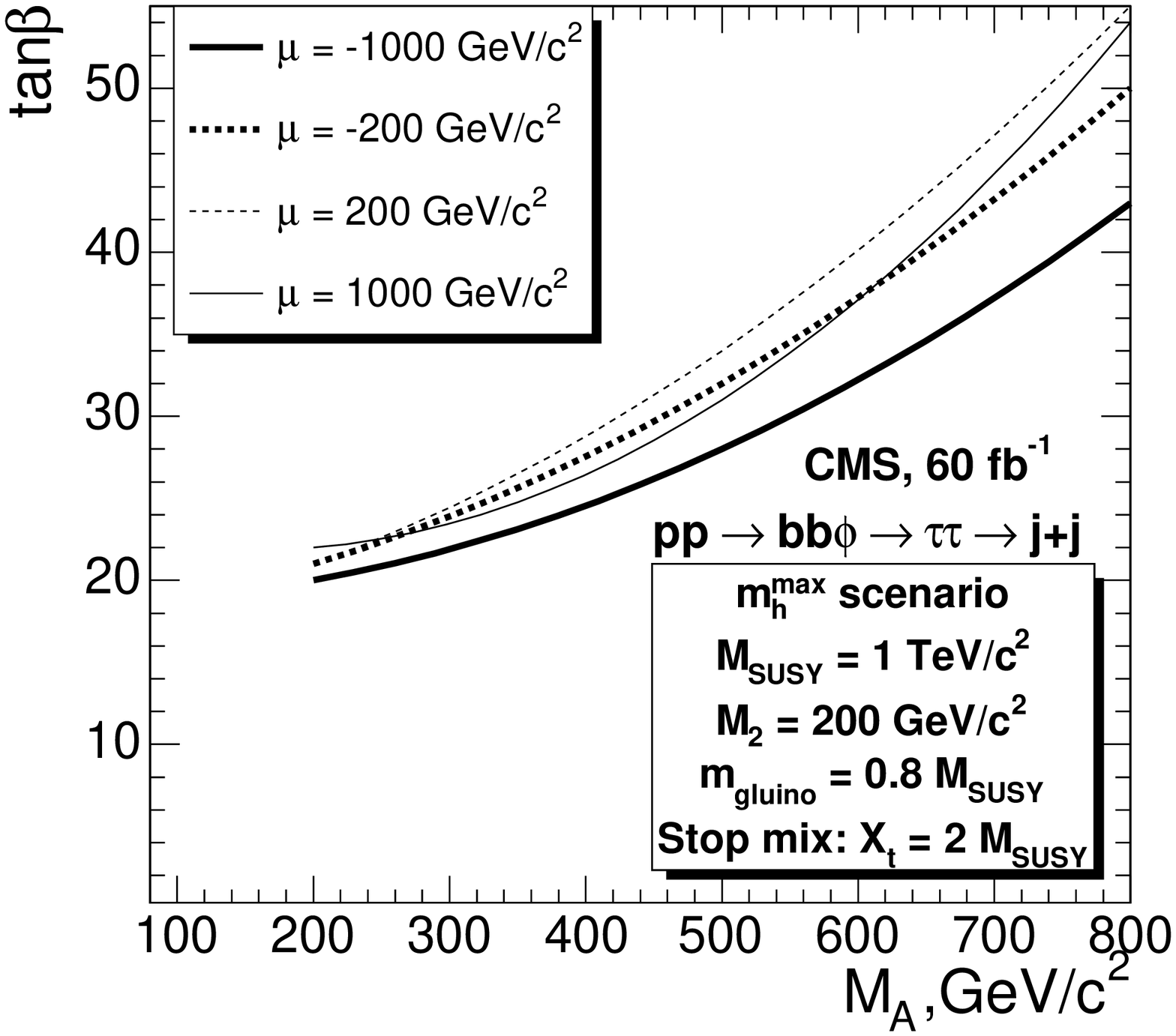}
\includegraphics[width=.45\textwidth]{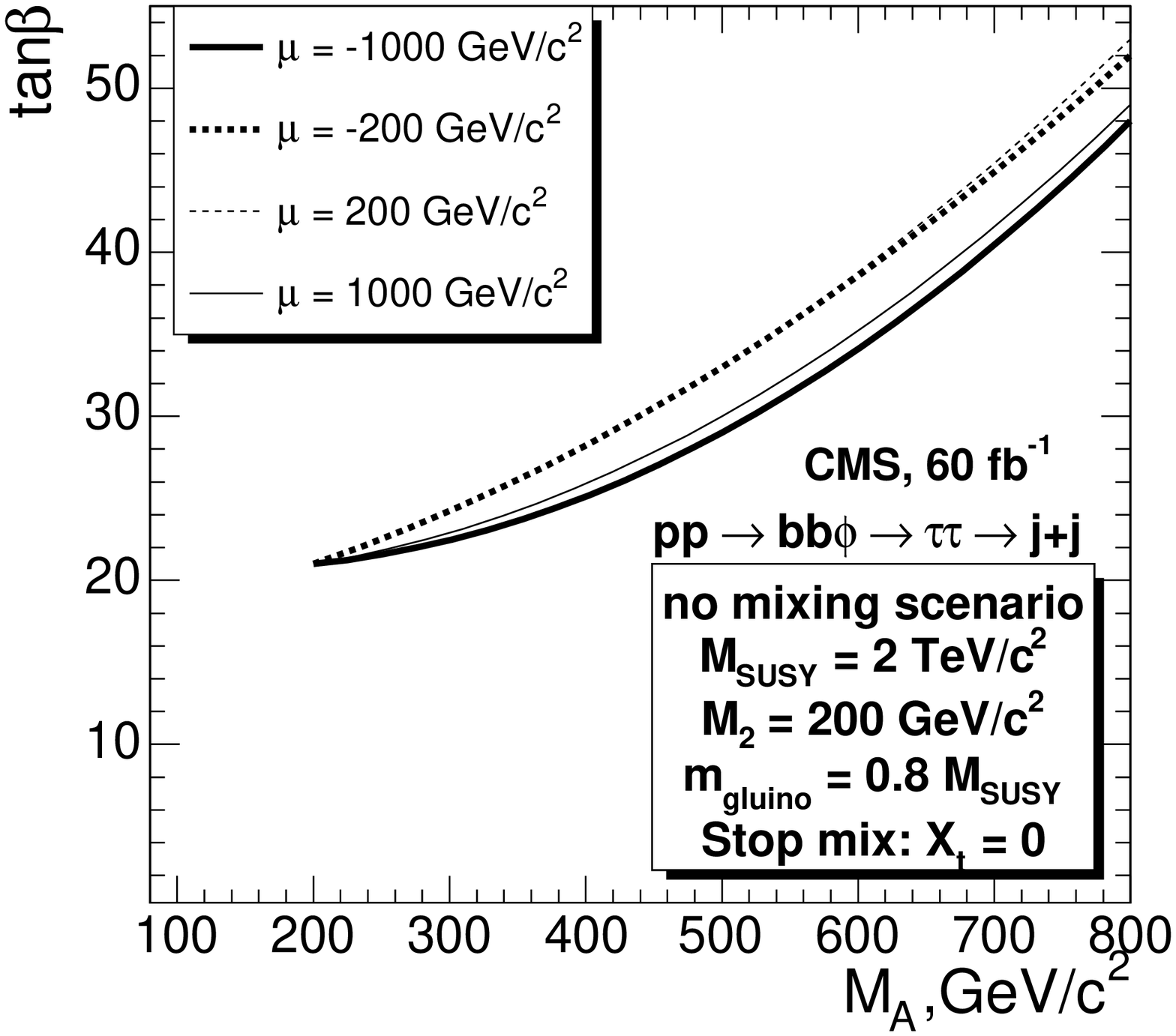}
\caption{%
The $5\,\si$ discovery regions (i.e.\ the upper bound of the ``LHC
wedge'' region) for the heavy neutral Higgs bosons in the channel 
$b \bar b \to b \bar b \; H/A , H/A \to \tau^+\tau^- \to \mbox{jets}$
(taken from \citere{cmsHiggs}).
}
\label{fig:LHCwedge}
\end{center}
\end{figure}




\end{document}


\end{document}
